\documentclass[structabstract]{aa}  
\usepackage{graphicx}
\usepackage{txfonts}
\usepackage{natbib}
\bibpunct{(}{)}{;}{a}{}{,}
\def\RRgband{0.1198$_{-0.0013}^{+0.0026}$}
\def\RRrband{0.1168$\pm$0.0010}
\def\RRiband{0.1162$\pm$0.0013}
\def\RRzband{0.1165$\pm$0.0013}
\def\RRriband{0.1143$\pm$0.0018}
\def\RRiiband{0.1162$\pm$0.0005}
\def\RRKsband{0.1189$\pm$0.0015}
\def\RRKcband{0.1162$_{-0.0021}^{+0.0016}$}
\def\RRKcbandb{0.1187$\pm$0.0018}
\def\RRKcbandc{0.1162$\pm$0.0030}

\def\Tcgband{55380.59049$\pm$0.00040}
\def\Tcrband{55380.59045$\pm$0.00019}
\def\Tciband{55380.59044$\pm$0.00022}
\def\Tczband{55380.59013$\pm$0.00017}
\def\Tcriband{55342.66081$\pm$0.00029}
\def\Tciiband{55407.45754$\pm$0.00007}
\def\TcKsband{55342.66057$\pm$0.00023}
\def\TcKcband{55426.42274$\pm$0.00032}

\def\T0{2455342.66068}
\def\sigT0{$\pm$0.00006}
\begin{document}

   \title{Optical to near-infrared transit observations of super-Earth GJ1214b: water-world or mini-Neptune?\thanks{Photometric timeseries are only available in electronic form at the CDS via anonymous ftp to cdsarc.u-strasbg.fr (130.79.128.5) or via http://cdsweb.u-strasbg.fr/cgi-bin/qcat?J/A+A/}}
   \author{E.J.W. de Mooij\inst{1}, 
           M. Brogi\inst{1},
           R.J. de Kok\inst{2},
           J. Koppenhoefer\inst{3,4}, 
           S.V. Nefs\inst{1},
           I.A.G. Snellen\inst{1},
           J. Greiner\inst{4}, 
           J. Hanse \inst{1},
           R.C. Heinsbroek\inst{1},
           C.H. Lee\inst{3},
           \and
           P.P. van der Werf\inst{1}}
   \institute{Leiden Observatory, Leiden University, Postbus 9513, 2300 RA, Leiden,     
               The Netherlands; \email{demooij@strw.leidenuniv.nl}
               \and SRON Utrecht, Utrecht, The Netherlands, 
               \and Universit\"ats-Sternwarte M\"unchen, Scheinerstrasse 1, 81679, Munich, Germany; 
               \and Max-Planck-Institut f\"ur extraterrestrische Physik, Giessenbachstrasse, 85748 Garching, Germany;  
              }  
   \date{  }
   \authorrunning{De Mooij et al.}
   \titlerunning{Transit observations of GJ1214b}

\abstract
{GJ1214b, the 6.55 Earth-mass transiting planet recently discovered by the MEarth team, has a mean density of $\sim$35\% of that of the Earth. It is thought that this planet is either a mini-Neptune, consisting of a rocky core with a thick, hydrogen-rich atmosphere, or a planet with a composition dominated by water.}
{ In the case of a hydrogen-rich atmosphere, molecular absorption and scattering processes may result in detectable radius variations as a function of wavelength. The aim of this paper is to measure these variations.}
{ We have obtained observations of the transit of GJ1214b in the r- and I-band with the Isaac Newton Telescope (INT), in the g-, r-, i- and z-bands with the 2.2 meter MPI/ESO telescope, in the K$_s$-band with the Nordic Optical Telescope (NOT), and in the K$_c$-band with the William Herschel Telescope (WHT). By comparing the transit depth between the the different bands, which is a measure for the planet-to-star size ratio, the  atmosphere is investigated.}
{ We do not detect clearly significant variations in the planet-to-star size ratio as function of wavelength. Although the ratio at the shortest measured wavelength, in g-band, is 2$\sigma$ larger than in the other bands. The uncertainties in the K$_s$ and K$_c$ bands are large, due to systematic features in the light curves.}
{The tentative increase in the planet-to-star size ratio at the shortest wavelength could be a sign of an increase in the effective planet-size due to Rayleigh scattering, which would require GJ1214b to have a hydrogen-rich atmosphere. If true, then the atmosphere has to have both clouds, to suppress planet-size variations at red optical wavelengths, as well as a sub-solar metallicity, to suppress strong molecular features in the near- and mid-infrared.
However, star spots, which are known to be present on the host-star's surface, can (partly) cancel out the expected variations in planet-to-star size ratio, because the lower surface temperature of the spots causes the effective size of the star to vary with wavelength.  A hypothetical spot-fraction of $\sim$10\%, corresponding to an average stellar dimming of $\sim$5\% in the i-band, would be able to raise the near- and mid-infrared points sufficiently with respect to the optical measurements to be inconsistent with a water-dominated atmosphere. Modulation of the spot fraction due to the stellar rotation would in such case cause the observed flux variations of GJ1214.}


\maketitle

\section{Introduction}

During the transit of an exoplanet, the light from its host-star filters through the planet's atmosphere, and atmospheric signatures from molecules and atoms get imprinted on the transmission spectrum. The strength of the atmospheric features is dependent on the atmospheric scale height H, which in turn depends on the temperature T, the mean molecular weight $\mu$ and the planet's surface gravity g as H=$\frac{kT}{\mu g}$. For hot-Jupiters this scale height is a few hundred kilometers. For instance, the expected atmospheric scale height of HD189733b for an H$_2$ dominated atmosphere is $\sim$200 km, while for HD209458b this is about 700 km. For transit observations, the observable parameter we are interested in is $\Delta$R$_p$/R$_*\sim$H/R$_*$, which is $\sim$0.0004 for HD189733b, and $\sim$0.0009 for HD209458b. Nevertheless, signatures from the atmospheres of hot-Jupiters have been detected in transmission spectra, both from atoms, including sodium~\citep[eg.][]{charbonneauetal02,snellenetal08,redfieldetal08}, potassium~\citep{singetal11a,colonetal11}, hydrogen~\citep{vidalmadjaretal03}, carbon~\citep{vidalmadjaretal04} and oxygen~\citep{vidalmadjaretal04}, as well as from molecules such as water~\citep{tinettietal07}, although this is disputed by \cite{desertetal09}, methane ~\citep{swainetal08}, which has been challenged by \cite{singetal09} and \cite{gibsonetal11},  and carbon monoxide~\citep{snellenetal10}. In addition, a gradual increase of the planet-to-star radius ratio of HD189733b has been detected toward shorter wavelengths, which has been attributed to the scattering by haze particles~\citep{pontetal08,singetal11b}.

For cooler and smaller planets, the scale height decreases, and the $\Delta$R$_p$/R$_*$ becomes much smaller. For the Earth the scale height is only 8.5~km, which corresponds to a change in the radius ratio of $\sim$10$^{-5}$, which is very small. However, for the recently discovered super-Earth GJ1214b~\citep{charbonneauetal09}, the scale height can be similar to that of HD189733b, due to the low ($\sim$0.35$\rho_{earth}$) density of the planet, if the atmosphere is dominated by H$_2$. Since its host-star, GJ1214, is $\sim$4 times smaller than the host-star of HD189733b, the change in the planet-to-star radius ratio is $\sim$4$\times$ larger at $\Delta$R$_p$/R$_*$=0.0016. This makes GJ1214b an ideal candidate to search for the signatures of its atmosphere, despite its small mass and radius.

\cite{rogersetal10} presented three formation scenarios for GJ1214b that explain its low density. These scenarios also provide predictions on the composition of the planets atmosphere. If GJ1214b formed as a predominantly water-rich planet, the atmospheric scale height would be small, since the mean molecular weight of water is high. However, if GJ1214b's low density is due to out-gassing from a rocky planet, or due to it being formed as a mini-Neptune, the atmosphere is expected to consist predominantly of hydrogen and helium, with a low mean molecular weight. Atmospheric models by~\cite{millerricciandfortney10} showed that it is possible to get detectable signatures from an atmosphere with a large scale height, especially from molecules in near-infrared, but also from the scattering of light in the optical part of the spectrum.

Recently, the first transit transmission results for the atmosphere of GJ1214b have been presented in the literature.~\cite{beanetal10} found that their spectroscopy in the z-band showed no sign for the presence of a thick, hydrogen-rich atmosphere, which argues for a water-rich atmosphere, something which is also consistent with the mid-infrared observations of~\cite{desertetal11}. These observations are in contrast with the results from~\cite{crolletal11}, who show that the transit in the K$_s$-band is deeper than the transit in the J-band, and therefore consistent with an atmosphere with a large scale-height and low mean-molecular weight.

In this paper we present the results for our multiband transit photometry of GJ1214b, covering a wavelength range from the g-band in the optical to the K$_c$-band in the near-infrared. In section~\ref{sec:obs} we present our observations, followed by the data reduction in section~\ref{sec:dr} and transit fitting in section~\ref{sec:tf}. Subsequently we discuss the influence of stellar variability in section~\ref{sec:starvar} and present and discuss the transmission spectrum of GJ1214b in section~\ref{sec:discuss}. Finally we give the conclusions in section~\ref{sec:concl}.

\section{Observations}\label{sec:obs}
\subsection{WFC observations}
\begin{figure}
\centering
\includegraphics[width=8.8cm]{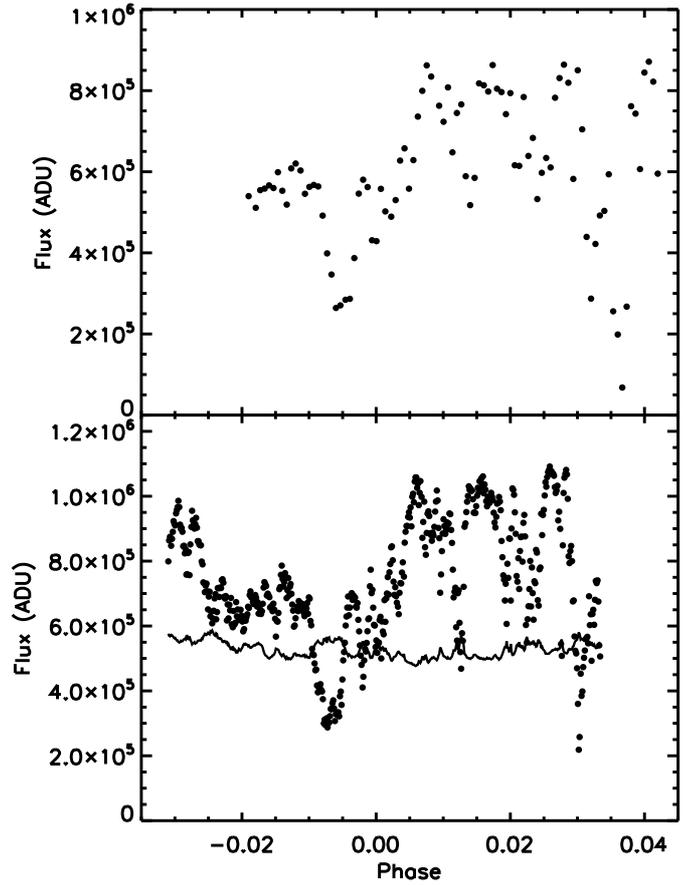}
\caption{Raw light curves for the INT r-band (top panel) and the NOT K$_s$-band observations (bottom panel). In the bottom panel we have also overplotted the flux in the aperture due to the sky-background}
\label{fig:LC_raw}
\end{figure}
A transit of GJ1214b was observed in Sloan r-band ($\lambda_c$=627~nm) on May 26, 2010  between 2:57 UT and 5:15 UT with the Wide Field Camera (WFC) on the 2.5 meter Isaac Newton Telescope (INT). An exposure time of 60 seconds was used resulting in 89 frames with an average cycle time of 93 seconds. Only the central detector of the WFC (CCD4) was used for the analysis, with a pixel scale of 0.33''/pixel, this CCD has a field of view of 675'' by 1350'', sufficient to observe a number of reference stars of similar brightness simultaneously with the target. 

The moon was almost full and the conditions were strongly non-photometric, with the transparency dropping to below 20\% during several frames (see top panel of  Fig.~\ref{fig:LC_raw}). 

On July 29, 2010, a transit of GJ1214 was observed in I-band ($\lambda_c$=822~nm) with the same instrument. The observations started at  21:23 UT and lasted for just over 3 hours. An exposure time of 50 seconds was used, resulting in a total of 142 frames with an average cycle time of 81 seconds. In this case the night was photometric.

Since GJ1214b is about 5.5 times brighter in I-band than in r-band, we significantly defocused the telescope in the I-band in order to keep the peak count-levels in the linear regime of the detector. This has the added benefit that the light is spread over more pixels, reducing the impact of flat-fielding errors.

\subsection{GROND griz-band observations}
On the night of July 3, 2010, we obtained simultaneous observations of GJ1214b in the g ($\lambda_c$=459~nm), r ($\lambda_c$=622~nm), i ($\lambda_c$=764~nm), and z-band ($\lambda_c$=899~nm) with the GROND instrument~\citep{greineretal08} on the 2.2 meter MPI/ESO telescope at La Silla in Chile. The field of view in each of the wavelength channels is 5.4' by 5.4', which is sufficient to observe both GJ1214 and a set of reference stars simultaneously. The observations started at 00:16 UT and lasted until 04:06 UT. During this time we obtained 280 frames in each of the four optical bands. The exposure time was varied from 20 to 30 seconds to avoid saturation of the CCDs. The average cycle time was 50 seconds. 

\subsection{NOTCam K$_s$-band observations}
We obtained a K$_s$ band transit observations with the NOTCam instrument on the Nordic Optical Telescope (NOT) simultaneously with our INT r-band observations on May 26, 2010.
The  observations were carried out in service mode, and the wide field imaging optics and a K$_s$-band filter ($\lambda_c$=2.15~$\mu$m) were used. The pixelscale of this setup is 0.234''/pixel, resulting in a field of view of the detector of 4 by 4 arcminutes. This field of view is sufficient to allow simultaneous observations of one reference star of similar brightness to GJ1214 as well as a reference star that is 4$\times$ fainter than GJ1214. The field of view of the detector was rotated to make sure that bad regions on the detector were avoided for all three stars. Since the NOT is located on the same mountain as the INT, these observations suffer from the same non-photometric conditions, with the transparency dropping to 25\% for parts of the light curve (see Fig.~\ref{fig:LC_raw}). This strongly affects the observations, since the sky background dominates over the object flux for the larger apertures, especially during times of low transparency.

Since GJ1214 is bright at near-infrared wavelengths, we defocused the telescope in order to allow for the relatively long integration time and reducing the sensitivity to flat-fielding errors, although this also increases the impact of the sky background.

The exposure time was set to 4 seconds, to allow for relatively efficient observations, with the large overheads induced by the NOTCam system. The average cycle time was 16 seconds, allowing us to capture 518 frames in 2 hours and 25 minutes of observations.

In order to increase the stability of the system for the observations, and to decrease the telescope overheads, we observed in staring mode, with guiding, keeping the centroid of the star constant to within 4 pixels during the observations. Since this observation strategy does not allow us to subtract the background from the images, we obtained a set of dithered observations after our transit observation, from which we constructed a background map.

\subsection{LIRIS K$_c$-band observations}
The K$_c$-band ($\lambda_c$=2.27$\mu$m) observations were obtained with the Long-slit Intermediate Resolution Infrared Spectrograph~\cite[LIRIS;][]{LIRIS02}  on the  William Herschel Telescope (WHT) on the night of August 17, 2010. The pixel-scale of the LIRIS detector is 0.25''/pixel, resulting in a field of view of the detector of 4.2 by 4.2 arcminutes. The  same reference stars as for the NOTCam observations were used.

A similar observing strategy as that of the NOTCam observations was followed, defocusing the telescope and performing the observations in staring mode. An exposure time of 10 seconds was used, resulting in an average cycle time of 12.5 seconds. We obtained 1010 frames during the 3 hours and 26 minutes of observations, of which 300 were taken during the transit. From these we discarded the first 40 frames, 28 of which were taken with the stars on a different position on the detector, and the others because they were taken with a different defocus of the telescope. The frames were obtained in sequences of 100, from which the first 3 frames clearly suffered from the reset anomaly. These frames were also excluded from further analysis (27 frames in total). As for the NOTCam observations, we observed a field offset from GJ1214 in order to construct a background map.

\section{Data reduction}~\label{sec:dr}
\subsection{Optical data}
\begin{figure*}
\centering
\includegraphics[width=16cm]{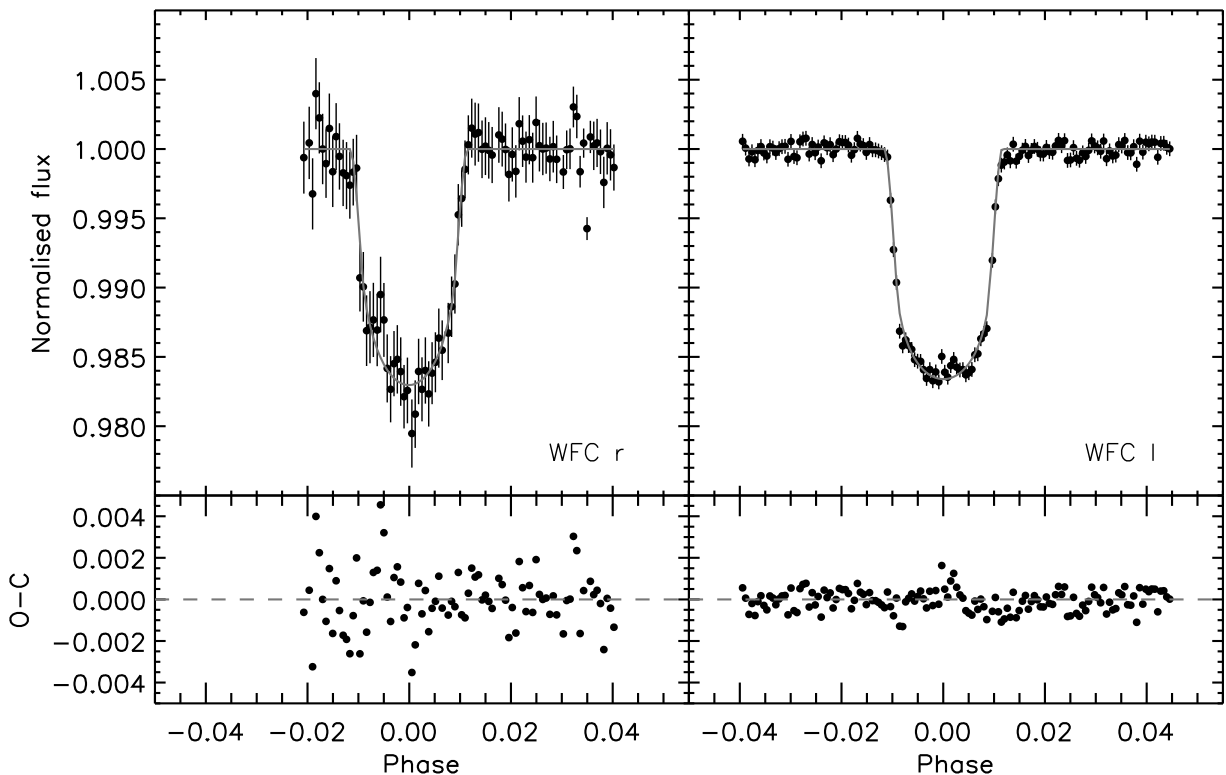}
\caption{Light curves for our WFC r- and I-band observations. The best fit transit models are overplotted. The r-band data were detrended on a point-by-point basis (see text), and both light curves have been corrected for airmass. The errorbars shown are set to the scatter in the out-of-transit baseline.}
\label{fig:LC_INT}
\end{figure*}
\begin{figure*}
\centering
\includegraphics[width=16cm]{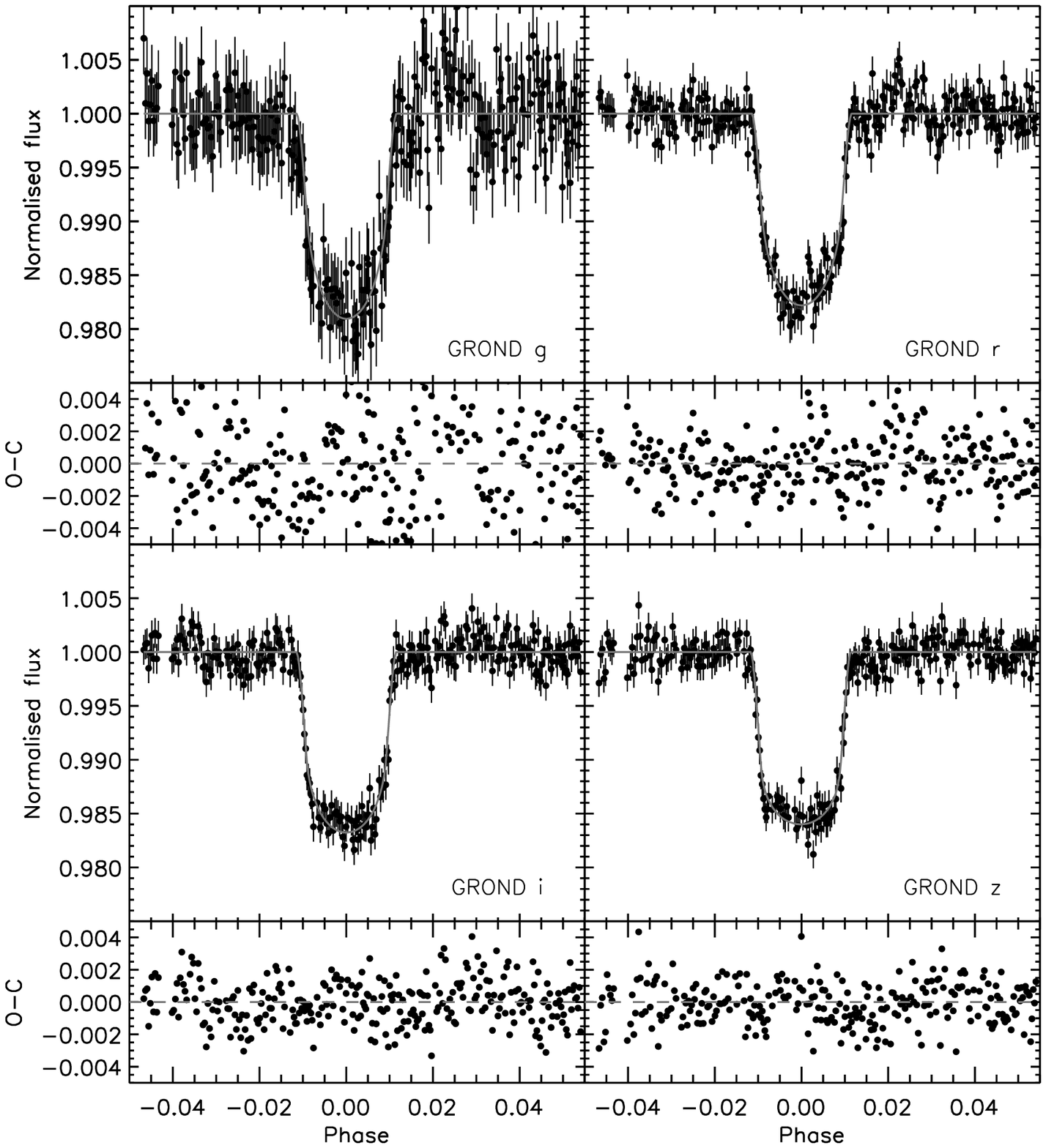}
\caption{The light curves for our GROND observations with the best fit models overplotted. All light curves have been corrected for a linear trend in the baseline. The errorbars shown are set to the scatter in the out-of-transit baseline.}
\label{fig:LC_GROND}
\end{figure*}
A standard data reduction was performed for the WFC-INT data: after bias subtraction the images were flatfielded with a twilight flat. In addition a fringemap was used to correct the WFC I-band data for strong fringing. 

Subsequently aperture photometry was performed on GJ1214 and a set of reference stars (24 for the r-band and 14 for the I-band), using apertures of 10 and 19 pixels for the r- and \mbox{I-band} respectively. The background was measured in annuli centered on the stars, using an inner radius of 25 and an outer radius of 50 pixels for the WFC r-band and 35 and 70 pixels for the I-band.

Since the weather conditions during the INT r-band observations were strongly non-photo-metric, extinction gradients could be present in the photometry of the stars across the detector. Rather than directly combining light curves of all stars to obtain a reference light curve, we fitted the normalised photometry of the reference stars in each frame with linear function in the (x,y)-position and background flux. We then calculated the value of the best fit function for GJ1214, to remove the extinction effects of the atmosphere. Subsequently we fitted an airmass curve to the corrected light curve of GJ1214 to adjust for second order colour effects. The resultant light curve is shown in the left panel of Fig.~{\ref{fig:LC_INT}}. The out-of-transit noise-level is 1.6$\cdot$10$^{-3}$ per data point, which is 1.3 times larger than expected from Poisson and read-noise statistics.

Since the WFC I-band data were taken under photometric conditions, we simply combined the photometry of all the reference stars for the reference light curve. Subsequently we corrected for the airmass as for the r-band data. The resultant light curve is shown in the right panel of Fig.~\ref{fig:LC_INT}. The out-of-transit noise-level is 4.5$\cdot$10$^{-4}$, while the expected noise level from Poisson and read-noise statistics is 3$\cdot$10$^{-4}$. However, this estimate does not take scintillation noise into account, which is difficult to quantify, and could easily dominate the noise at this level.

The GROND data, taken in the g-, r-, i- and z-bands simultaneously, were reduced in a similar way as the INT data. An aperture of 18 pixels was used for the aperture photometry, while the background was determined in an annulus around the star with an inner radius of 20 pixels and an outer radius of 30 pixels for all bands. 

Subsequently, a linear base-line was fitted to the light curves of each band resulting in an out-of-transit RMS of the light curves of 3.3$\cdot$10$^{-3}$, 1.6$\cdot$10$^{-3}$, 1.4$\cdot$10$^{-3}$ and 1.3$\cdot$10$^{-3}$, for the g-, r-, i- and z-bands respectively. These noise-levels are significantly higher than expected from Poisson and read-noise statistics, possibly due to the fact that these in-focus observations are much more sensitive to flatfielding inaccuracies.

\subsection{Near-infrared data}
\begin{figure*}
\centering
\includegraphics[width=16cm]{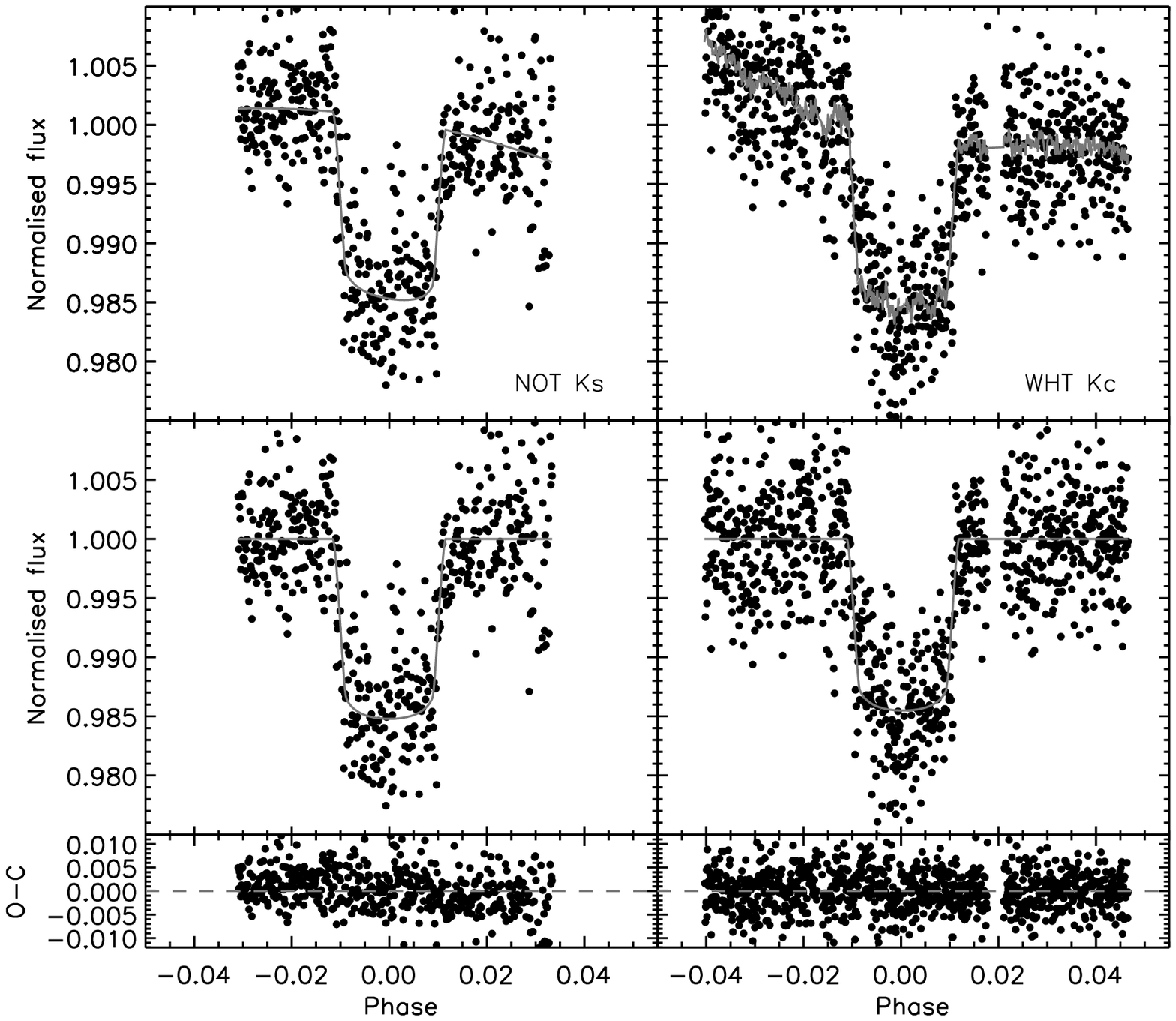}
\caption{Light curves for our NOTCam K$_s$-band and LIRIS K$_c$-band observations. Top panels: Raw, normalised light curves with best fit model for the transit and systematic effects overplotted. Middle panels: Light curves corrected for systematic effects, with the best fit transit model overplotted. Bottom panel: residuals after subtracting the best fit models.}
\label{fig:LC_NIR}
\end{figure*}
As a first step in the NOTCam K$_s$-band data reduction a non-linearity correction was made to all frames on a pixel-by-pixel basis, using a non-linearity measurement from the NOTCam calibration archive~\footnote{http://www.not.iac.es/instruments/notcam/calibration.html}. The science frames were subsequently flatfielded using a twilight flat. Those frames obtained for sky-subtraction purposes were flat-fielded, and combined after removing the outliers for each pixel, which removes any stars present in the individual images. The result was subsequently subtracted from the science images, removing most of the structure in the background.

In the next step, aperture photometry was performed on GJ1214 and the reference stars, using an aperture of 10 pixels for the NOTCam dataset. The residual background  was determined by averaging the background signal in an annulus centered on the stars with an inner radius of 30, and an outer radius of 50 pixels. Individual pixels in the annulus were clipped at 5-$\sigma$ from the mean to prevent outliers from affecting the background measurements. 

The WHT-LIRIS K$_c$-band data were reduced in a similar way, but we constructed our own non-linearity curve from a sequence of dome-flats at different exposure times. After the sky-subtraction, aperture photometry was performed with a 25 pixel radius, using an annulus from 30 to 50 pixels to determine the residual background, clipping the outliers in the same way as for the NOTCam data. 

The NOTCam K$_s$- and LIRIS K$_c$-band light curves normalised by those from the reference stars are shown in the top-panels of Fig.~\ref{fig:LC_NIR}. Further corrections for systematic effects are done simultaneously with the transit fitting and are discussed in Sect.~\ref{sec:tf_nir}.

\section{Transit fitting}\label{sec:tf}
\subsection{Optical transits}
Since our goal is to measure the effective radius of the planet as a function of wavelength and not to determine the other transit parameters, we fit the transit at each band independently, keeping the impact parameter and semi-major axis fixed to the values used by~\cite{beanetal10}. This has the added benefit that we can compare our results directly to previous literature values. To model the transit curve we use the formalism of~\cite{mandelandagol02} with non-linear limb-darkening coefficients as calculated by~\cite{claret00} for the INT I-band observations and by~\cite{claret04} for the observations at the other wavelengths. We used the limb-darkening parameters for the gridpoint closest to the stellar parameters from~\cite{charbonneauetal09}.  The limb-darkening coefficients are given in Table~\ref{tab:ldc}. We performed a Markov-Chain Monte Carlo (MCMC) analysis to find the planet-to-star radius ratio and the time of mid-transit, creating 5 chains with a length of 200,000 points for each band. From each of the chains the first 20,000 points were trimmed to ensure that the initial conditions did not influence the results. After checking that the chains were well mixed using the~\cite{gelmanandrubin92} test, we merged them. The best fit models are overplotted on the light curves in Figs.~\ref{fig:LC_INT}~and~\ref{fig:LC_GROND}. 

To assess the impact of red-noise on uncertainties of the fitted parameters, we used the residual permutation method, which uses the data itself to assess the uncertainties on the measured parameters due to correlated noise. The best fit model is added to the residuals from the MCMC fit, after the residuals were shifted by N points. The points are wrapped around when performing the shift. The new light curve is then refitted. For the fitting we used the IDL MPFIT package\footnote{http://purl.com/net/mpfit}~\citep{markwardt09}. To assess the uncertainties on the parameters we use the range between 16\% and 84\% of the distribution of the parameters for the $\pm$1-$\sigma$ uncertainty interval. In the INT r- and I-band observations, the uncertainty estimates from the residual permutation method are lower than those obtained from the MCMC analysis, this is probably due to the low number of frames that were obtained. In this case we have adopted the uncertainties from the MCMC analysis rather than those from the residual permutation analysis.

The fitted radius ratios for all bands  (see also Table~\ref{tab:rratios}), are \RRriband\ and \\ \RRiiband, for our INT r and I-band measurements, and \RRgband, \RRrband, \RRiband\ and \RRzband\ for our GROND g-, r-, i- and z-band observations respectively. The times of mid-transit are given in Table~\ref{tab:fitpars}.

\subsection{Near-infrared transits}\label{sec:tf_nir}
\begin{figure}
\centering
\includegraphics[width=8.8cm]{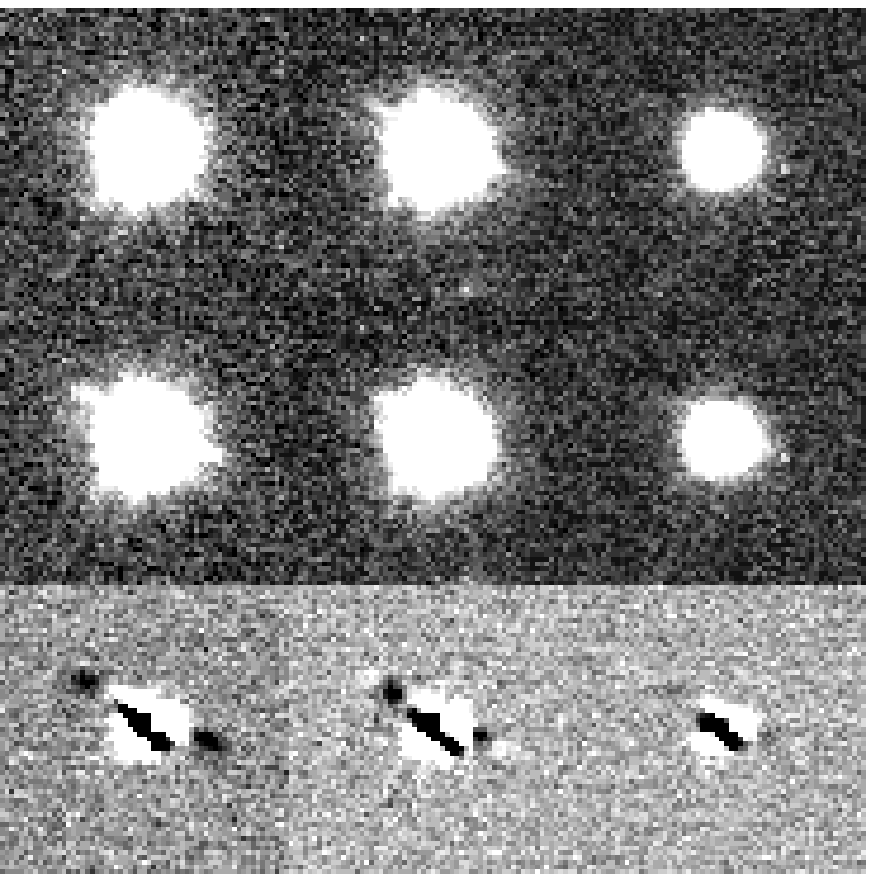}
\caption{Images of the PSFs for our LIRIS K$_c$-band observations taken at two different times. The left column is for GJ1214, while the middle and right columns are for the two reference stars. The top panels show the average PSF in a sequence of 31 frames just before an anomalous feature is visible in the light curve, while the middle row shows the average PSF during the anomalous feature. The difference between the PSFs is shown in the bottom row.}
\label{fig:PSFcompKc}
\end{figure}
Firstly, outliers were removed from the K$_s$-band light curve by excluding all points more than 1.2\% away from a median smoothed light curve. In this way 25 points were removed. A clear residual trend in the out-of-transit baseline is visible in the light curve (Fig.~\ref{fig:LC_NIR}), but we find no significant correlations with the position on the detector or airmass, which we often see in other near-infrared measurements~\citep[e.g.][]{demooijandsnellen09,demooijetal11}. We therefore fitted the light curve with a second order polynomial simultaneously with the transit parameters. As for the optical light curves, we only fitted for the time of mid-transit and the planet-to-star size ratio, keeping the impact parameter and semi-major axis fixed to the values used by~\cite{beanetal10}. The fits were again performed using an MCMC analysis, as for the optical data, using 5 chains of 200,000 steps, trimming away the first 20,000 points and checking that they were well mixed~\citep{gelmanandrubin92}. Non-linear limb-darkening parameters from~\cite{claret00} were used (given in Table~\ref{tab:ldc}). The best fit model is overplotted in the top-left panel of Fig.~\ref{fig:LC_NIR}. The light curve with the corrected baseline is shown in the left middle panel, and the residuals in the bottom left panel of the same figure.

Just as for the optical data, the impact of correlated noise is investigated using the residual permutation method. The estimates for the uncertainties given by the residual permutation method are almost identical to those we obtain from our MCMC analysis, which indicates that, apart from the trend in the baseline, the systematic effects are small compared to the random noise. We measure a radius ratio in the K$_s$-band of \RRKsband.

The LIRIS K$_c$-band light curves were fitted in a similar way to the NOTCam K$_s$-band data. First, 13 outlier points were removed from the light curve. In addition to the systematic trends seen in the baseline, which again do not correlate with airmass or with position on the detector, we found a correlation with the fraction of the total flux of the star contained in the brightest 11 pixels, which we take to be a proxy for the seeing (note that due to the strong defocus we cannot measure the seeing directly). We therefore fitted the relevant transit parameters simultaneously with a second order polynomial and this seeing proxy. We used a MCMC analysis similar to that used for the K$_s$-band light curve. In Fig.~\ref{fig:LC_NIR} the best fit model is overplotted in the top right panel, and the light curve corrected for systematic effects is shown in the middle right panel.
The residuals from the best fit model are shown in the bottom right panel of the same figure. 

The estimates for the uncertainties given by the residual permutation method are $\sim$50\% higher than those that were obtained from the MCMC analysis, and there is also an asymmetry between the upper and lower uncertainties.  

As can be seen in the bottom right panel of Fig.~\ref{fig:LC_NIR}, a significant feature is visible in the corrected K$_c$-band light curve between phase -0.012 and phase -0.006. Although this feature could be due to starspots, at the same moment a sudden increase in the difference in background occurs between the target and the reference star. At the moment of this  'bump' we also see a change in the shape of the PSF at lower intensities across the field of view of the detector (especially around the diffraction spikes caused by the support of the secondary mirror, see Fig.~\ref{fig:PSFcompKc}). It seems that instrument-related effects influence the measurements at this point. To check the influence of this feature on the fit to the light curve, we excluded it and refitted the time-series. While for the light curve as a whole we find a radius ratio of \RRKcband, we find a radius-ratio of \RRKcbandb\ ($\sim$1.5$\sigma$ deeper) when this feature is excluded from the fit. Since we do not want to bias our results, we use the radius-ratio obtained from fitting the entire K$_c$-band light curve for the rest of the paper. The extra uncertainty is accounted for by increasing the error on the radius ratio by 50\%.

\begin{table*}
\caption{Parameters used for the transit fitting of GJ1214 and the measured times of mid-transit. }
\label{tab:fitpars}
\centering
\renewcommand{\arraystretch}{1.35}
\begin{tabular}{l l c c}
\hline\hline
 Parameter & & Value &  Reference \\
\hline
a/R$_*$                  &        & 14.9749        & (1) \\
b=a/R$_*$cos(i)          &        & 0.27729        & (1) \\  
P                  & \multicolumn{1}{c}{(days)} & 1.580408346  & (1) \\  
T$_{star}$                & \multicolumn{1}{c}{(K)}    & 3030           & (2) \\
\hline                   
T$_C$ -2,400,000  (HJD)       & (INT r)     & \Tcriband      & (3) \\
\multicolumn{1}{c}{\ldots}    & (NOT K$_s$) & \TcKsband      & (3) \\
\multicolumn{1}{c}{\ldots}    & (GROND g)   & \Tcgband       & (3) \\
\multicolumn{1}{c}{\ldots}    & (GROND r)   & \Tcrband       & (3) \\
\multicolumn{1}{c}{\ldots}    & (GROND i)   & \Tciband       & (3) \\
\multicolumn{1}{c}{\ldots}    & (GROND z)   & \Tczband       & (3) \\
\multicolumn{1}{c}{\ldots}    & (INT I)     & \Tciiband      & (3) \\
\multicolumn{1}{c}{\ldots}    & (WHT K$_c$) & \TcKcband      & (3) \\
\hline
\end{tabular}
\tablebib{
(1)~\cite{beanetal10}; (2)~\cite{charbonneauetal09};  (3) this work.
}
\end{table*}
\begin{table*}
\caption{ Limb-darkening coefficients used for the fitting of the transit light curves. }
\label{tab:ldc}
\centering
\renewcommand{\arraystretch}{1.35}
\begin{tabular}{l c c c c c r}
\hline
\hline
Instrument &  Band & C$_1$ & C$_1$ & C$_1$ & C$_1$ & Reference  \\
\hline
GROND  & g     &0.2306 &  1.0639 & -0.3180 &  0.0119 & (2)\\
GROND  & r     &0.5654 &  0.1274 &  0.5991 & -0.3119 & (2)\\
WFC    & r     &0.5654 &  0.1274 &  0.5991 & -0.3119 & (2)\\
GROND  & i     &0.6819 &  0.4526 & -0.1692 & -0.0041 & (2)\\
WFC    & I     &0.6061 &  0.8676 & -0.7015 &  0.1881 & (1)\\
GROND  & z     &0.9613 &  0.5228 & -0.8702 &  0.3438 & (2)\\
NOTCam & K$_s$ &1.9371 & -2.8039 &  2.3942 & -0.8098 & (1)\\
LIRIS  & K$_c$ &1.9371 & -2.8039 &  2.3942 & -0.8098 & (1)\\
\hline
\end{tabular}
\tablebib{
(1)~\cite{claret00}; (2)~\cite{claret04}.
}
\end{table*}
\begin{table*}
\caption{Radius ratios from our observations} 
\label{tab:rratios}
\centering
\renewcommand{\arraystretch}{1.35}
\begin{tabular}{l c c c c r}
\hline
\hline
Instrument &  Band & $\lambda_c$  & $\left(\frac{R_p}{R_*}\right)_{measured}$ &$\Delta\left(\frac{R_p}{R_*}\right)_{corr}$ &$\Delta\left(\frac{R_p}{R_*}\right)_{10\% spots}$ \\
   &  &  ($\mu$m) &  &  \\
\hline
GROND  & g     & 0.46 & \RRgband   &   -    &  -0.0007 \\
GROND  & r     & 0.62 & \RRrband   &   -    &  -0.0005 \\
WFC    & r     & 0.63 & \RRriband  & 0.0011 &  -0.0006 \\
GROND  & i     & 0.76 & \RRiband   &   -    &   0.0000  \\
WFC    & I     & 0.82 & \RRiiband  &   -    &  +0.0005 \\
GROND  & z     & 0.90 & \RRzband   &   -    &  +0.0009 \\
NOTCam & K$_s$ & 2.15 & \RRKsband  & 0.0003 &  +0.0023 \\
LIRIS  & K$_c$ & 2.27 & \RRKcbandc &   -    &  +0.0021 \\
\hline
\end{tabular}
\tablefoot{Radius ratios as determined from our observations in different filters (column 4), the correction that is applied to the radius ratios due to stellar variability (column 5), as well as the wavelength-depedent change in radius ratio assuming a background of unocculted starspots that cover 10\% of the stellar surface (column 6) (note that the difference in i-band is set to 0) (see also Sect.~\ref{sec:spotbg}). The instruments, filters and central wavelengths are indicated in columns 1 through 3.
}
\end{table*}

\section{Stellar variability}\label{sec:starvar}
\cite{charbonneauetal09} found that the star GJ1214 is variable at the level of 1\% in the MEarth bandpass, with an apparent period of $\sim$80 days, attributed to the rotation-modulated variations in the star-spot levels. Recently~\cite{bertaetal11} presented a stellar variability analysis covering several years of data, finding a period of 52.3$\pm$5.3 days. This star-spot induced variability can have a major impact on transit photometry taken at different epochs and/or at different wavelengths~\citep[e.g.][]{carteretal11,singetal11b}. Since our data are taken over the course of 3.5 months, and our aim is to measure differences in the transit-depth as a function of wavelength, both the temporal and spectral effects from starspots are present and need to be taken into account. 

If a star-spot is present on the stellar surface during transit, firstly the planet could transit the spot. Because the star-spots have a lower surface brightness than the unspotted stellar surface, a spot that is occulted by the planet would make the transit less deep and bias the planet-to-star radius ratio to lower values. 
In the case the planet does not cross the star spot, the transit appears deeper, since the average surface brightness along the path of the planet is higher than that of the spotted surface, leading to a decrease of the effective radius of the star. Unocculted spots will therefore bias the planet-to-star size ratio to higher values. In addition, since the star spots have a lower effective temperature than the stellar surface, and are therefore relatively red, these effects will be stronger in the blue part of the optical spectrum than in the near-infrared. An expression for the bias due to unocculted starspots can be found in~\cite{singetal11b}, for more generalised equations for these corrections we refer the reader to \cite{desertetal2011}.

\subsection{Correcting for the stellar variability}\label{sec:starvarcor}
Changes in the spot fraction can be monitored by measuring the stellar brightness as a function of time, which can subsequently be used to link the observations taken at different epochs. Note that this does not take into account the base-level of spots, which is unknown but can still significantly influence the measured planet-to-star radius ratio as a function of wavelength. This is discussed in Sect.~\ref{sec:spotbg}.

To link our INT I-band observations to our INT r-band observations, we obtained a set of r-band out-of-transit frames on July 29, 2010, in order to measure the change in  stellar brightness between the two nights. These r-band frames were reduced in the same way as the transit observations, and aperture photometry was performed on the same set of stars as for the observations on May 26. We find that between May 26  and July 29 the star has decreased its brightness by $\Delta$F=2\%. We therefore correct our INT r-band and NOTCam K$_s$-band points to match the stellar brightness of the INT I-band observations. We use the expression for the effects of starspots on the planet-to-star radius ratio from~\cite{singetal11b}, where we use the NextGen models~\citep{hauschildtetal99} for flux of both the star and the spots. We assume a spot temperature of T$_{spot}$=2800~K, which is 200~K cooler than the effective temperature of the star. We find a correction $\Delta$R$_p$/R$_*$=0.0011 in the r-band and $\Delta$R$_p$/R$_*$=0.0003 in the K$_s$-band.

We linked the stellar brightness during the night of the GROND observations to that during the INT r-band observations using the data from~\cite{bertaetal11}. Subsequently, this was linked to our INT I-band observations using the offset found above. These two steps cancel each other out within the uncertainties and therefore no correction was performed on the GROND measurements. We also did not correct the K$_c$-band observation, since we have no flux measurements of the stellar brightness in the r-band on that night. However, the correction is expected to be small ($\Delta$R$_p$/R$_*\lesssim$0.0003), and will therefore not significantly influence the measured transmission spectrum. The corrections for the radius ratios are given in Table~\ref{tab:rratios}.

\section{Discussion}\label{sec:discuss}
\subsection{The transmission spectrum of GJ1214b}\label{sec:tspec}
\begin{figure*}
\centering
\includegraphics[width=16cm]{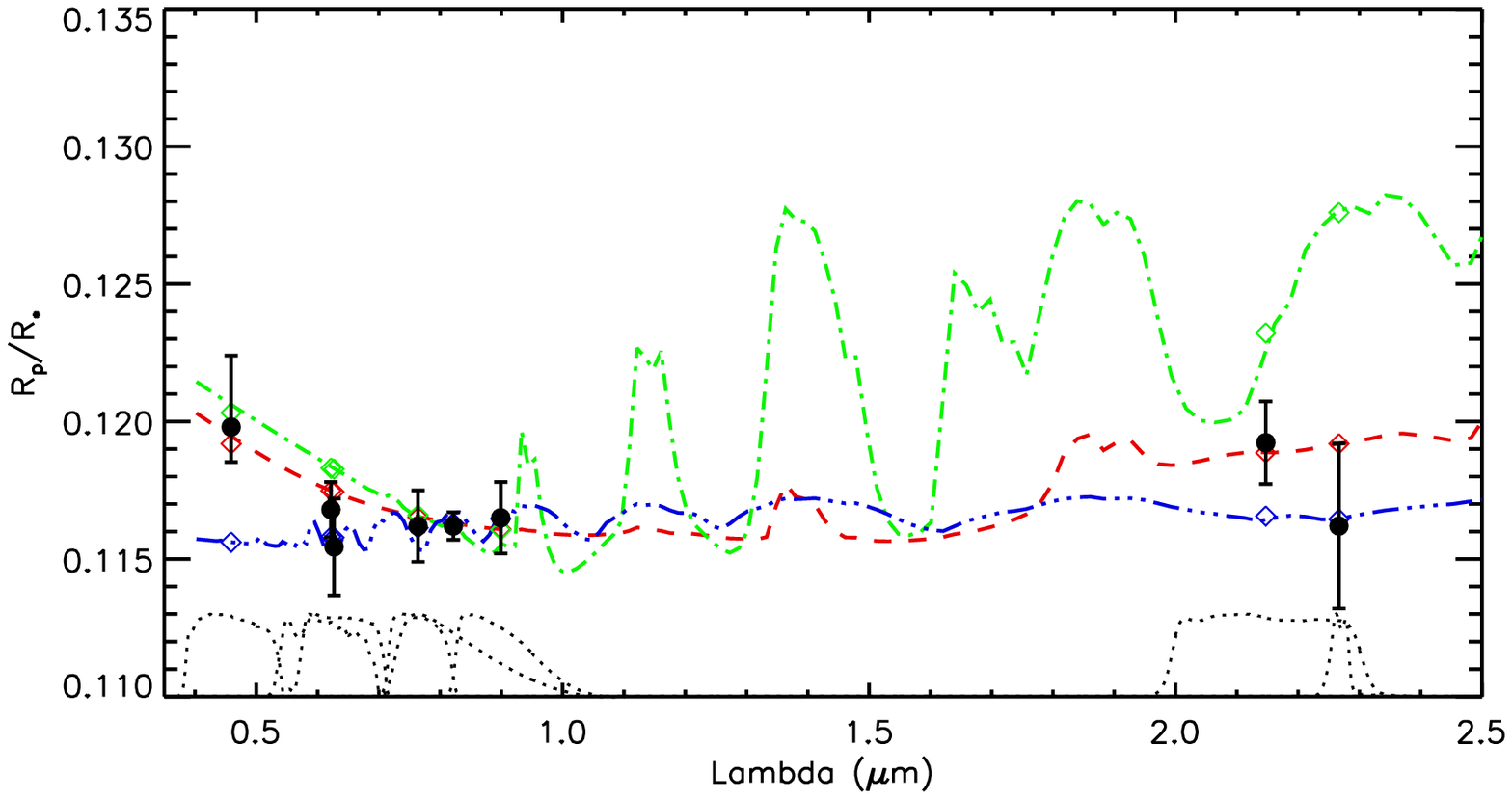}
\caption{Transmission spectrum of GJ1214b from our data. Overplotted are models for the atmosphere of GJ1214b. The green (dashed-dotted) line is a hydrogen rich atmosphere with a solar metallicity, the red (dashed) curve is for an atmosphere that is hydrogen rich with a sub-solar metallicity and a cloud layer at 0.5~bar, the blue (dashed-triple dotted) line is for an atmosphere that is dominated by water. The dotted curves at the bottom of the plot show the various transmission curves of the filters used for our measurements. }
\label{fig:tspec0}
\end{figure*}
Using the measured planet-to-star radius ratios in all of our bands, corrected for stellar variability, we can construct a transmission spectrum for GJ1214b, which is shown in Fig.~\ref{fig:tspec0}. All data points are consistent with a constant planet-to-star size ratio within 1$\sigma$, except for the g-band point at 0.46$\mu$m, which deviates by $\sim$2$\sigma$.

\subsection{Atmospheric models}
To investigate what constraints our observations can place on the atmosphere of GJ1214b, we compare our results with theoretical models of the planet's atmosphere. The models include Rayleigh scattering as well as molecular absorption features and collision-induced absorption from H$_2$-H$_2$~\citep{borysow02}. For the transmission spectrum we calculated the light passing through layers of atmosphere between 1$\cdot$10$^{-6}$ and 5 bar. At higher pressures we consider the atmosphere to be fully opaque. At each point we assume that the atmosphere can be described by an average profile in hydrostatic equilibrium. For the temperature-pressure profile we use one that is similar to that used by~\cite{millerricciandfortney10}. Our models include absorption by water, for which the opacity data from the HITEMP database~\citep{rothmanetal10} were used. In addition we also include methane, for which the data from HITRAN 2008~\citep{rothmanetal09} were used. Water and methane are expected to be the main absorbing gases in the atmosphere. A line-by-line code was used to calculate the opacities for the different gases, assuming a homogeneous mixing of all the different species present in the atmosphere. We used a Voigt profile for the individual lines and applied a line wing cut-off of 30 cm$^{-1}$ to simulate a sub-Lorentzian line profile \citep[e.g.][]{baileyandkedziora11}. This gives rise to a lower continuum and more pronounced absorption features than GJ1214b model results presented elsewhere.

We have chosen to generate three models to make a qualitative study of the atmosphere of GJ1214b, and we have overplotted these models on the observed transmission spectrum shown in Fig.~\ref{fig:tspec0}. The models were matched to the mean of the measured radius ratios between 0.7 $\mu$m and 1 $\mu$m. The first model, the green (dash-dotted) line in Fig.~\ref{fig:tspec0}, is for a (geometrically) thick atmosphere with a solar composition and no cloud layers. The concentrations of water and methane are 3$\cdot$10$^{-4}$. The dominant hydrogen gives rise to a high atmospheric scale height, as can be seen by the strong molecular features in the near infrared. Fig.~\ref{fig:tspec0} indicates that a solar metallicity atmosphere gives features that are too strong compared to our measurements, resulting in a $\chi^2$ of 26.7.
The second model, the red (dashed) line in the figure, is for an atmosphere with a sub-solar metallicity, and includes a grey cloud layer at a pressure of 0.5~bar. The concentrations of water and methane are 5.6$\cdot$10$^{-6}$ and 5$\cdot$10$^{-7}$ respectively. This model is consistent with our measurements ($\chi^2$=3.2). The third model, the blue (dashed-triple dotted) line in the figure is for a water dominated atmosphere. This model has its molecular features significantly suppressed due to the decreased scale height. This model is also consistent with all our observations, but with a higher $\chi^2$ of 15.1 compared to that of the second model. This is mainly due to the g-band measurement. The latter two models are chosen such that they can reproduce the relatively flat transmission profile seen in the red part of the optical spectrum. These $\chi^2$ values can be compared to that of a straight line, which has $\chi^2$=10.5.

The data can be well fitted with a model with a sub-solar metallicity and a cloud layer. The major disagreement between our data and the model for a water-dominated atmosphere comes from the GROND g-band measurements, which could be due to Rayleigh scattering, and to a lesser extent from the NOTCam K$_s$-band observations. It is therefore important to obtain further observations to confirm these measurements.

\subsection{Comparison with previous measurements}
\begin{figure*}
\centering
\includegraphics[width=16cm]{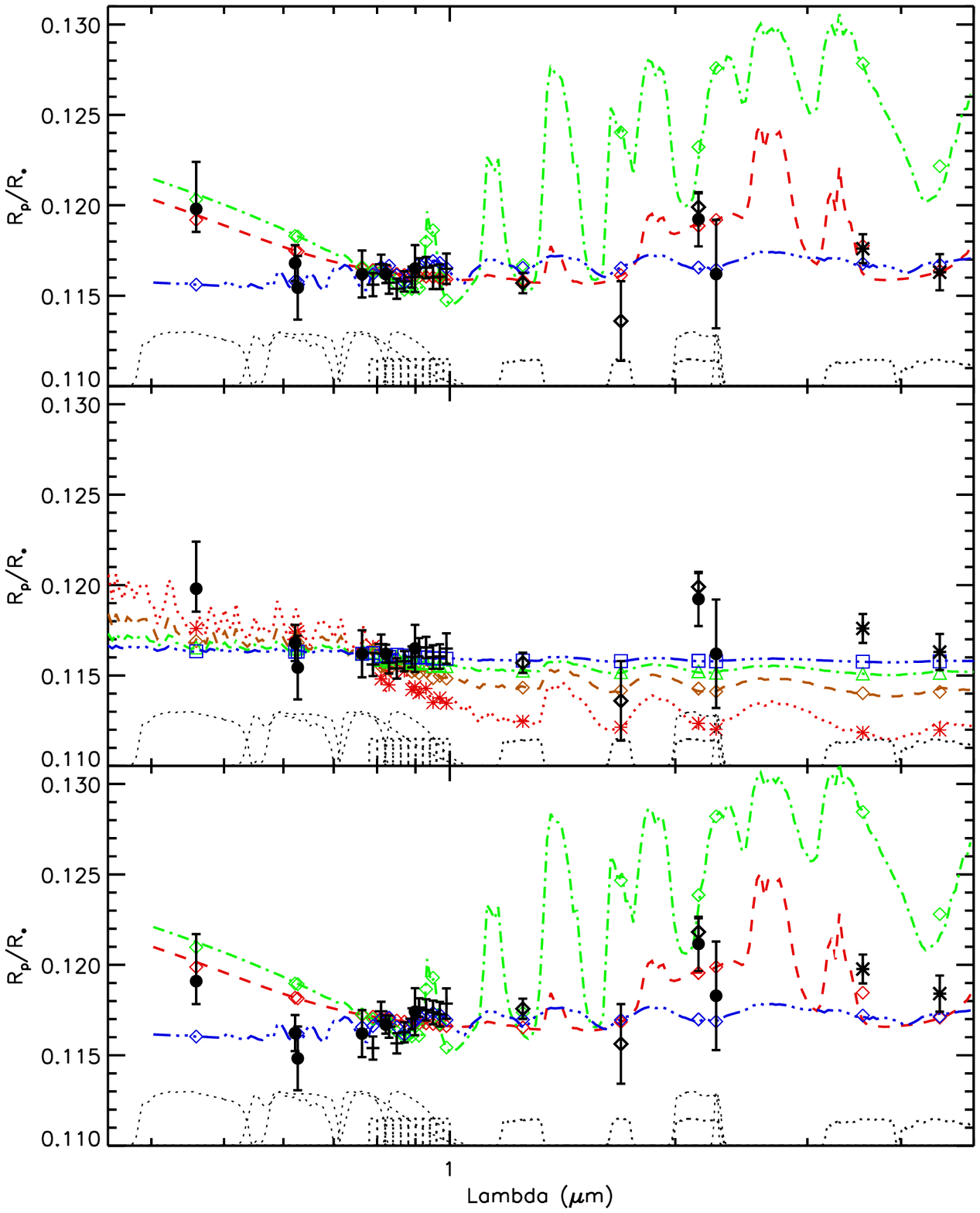}
\caption{Top panel: Transmission spectrum of GJ1214b. Overplotted are the atmospheric models for GJ1214, using the same linestyles as in Fig.~\ref{fig:tspec0} for the atmospheric models. Our data are plotted as filled circles, the VLT data from~\cite{beanetal10} as crosses, the {\it Spitzer} data from~\cite{desertetal11} as stars, and the data from~\cite{crolletal11} as diamonds. 
 Middle panel: The same data points as in the top-panel but now with different models overplotted showing the change in radius ratio due to different amounts for the base levels of stellar spots in the GROND i-band. The dotted, dashed, dash-dotted and the dashed-triple dotted lines are for 10\%, 5\%, 2.5\% and 1\% contribution from stellar spots in the i-band.
 Bottom panel: The observed transmission spectrum, but now corrected for a spot-dimming in i-band due to spots of 5\% (The dashed curve in the middle panel), which corresponds to a spot coverage of 10\% of the stellar surface.  The i-band radius ratio was kept at the same level as in the other panels. }
\label{fig:tspec_all}
\end{figure*}
\begin{figure*}
\centering
\includegraphics[width=16cm]{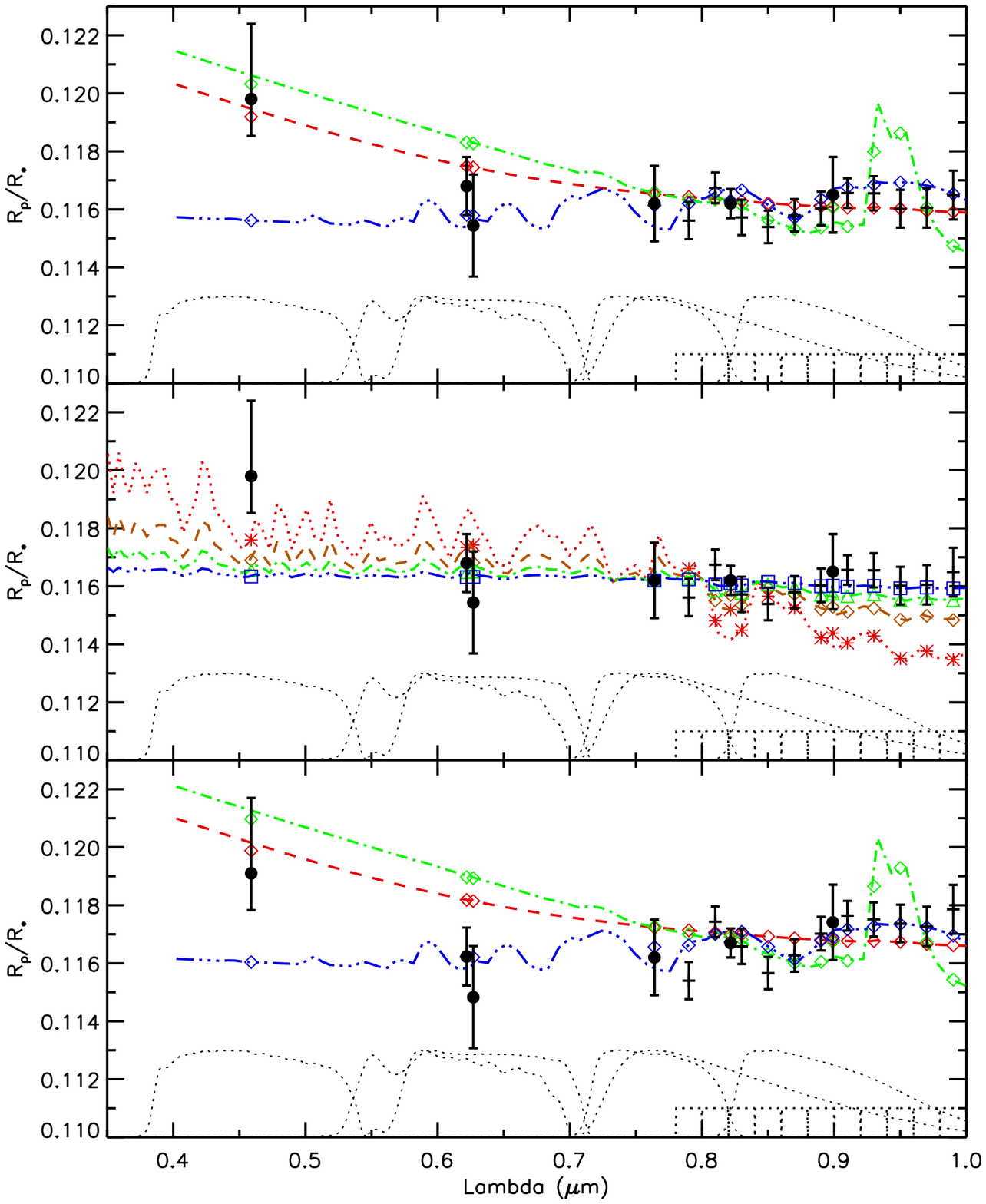}
\caption{Same as Fig.~\ref{fig:tspec_all} but now only showing the optical part of the transmission spectrum.}
\label{fig:tspec_all_opt}
\end{figure*}
During the past year there have been many transit measurements of GJ1214b at different wavelengths. \cite{beanetal10} measured the transit-depth in 11 wavelength bins  from 790~nm to 1000~nm. Our radius ratios obtained in the same wavelength range are consistent with their measurements. \cite{beanetal10} cannot explain their observations with a thick, hydrogen-rich atmosphere, and therefore favour models for a water-dominated atmosphere of GJ1214b. However, they cannot rule out clouds as an explanation for their relatively featureless transmission spectrum. This is also seen when comparing their data with our models, the $\chi^2$ for the water model is 9.1, while that for the model with a sub-solar composition with a cloud layer is 7.8, and for a straight line $\chi^2$=6.0.

\cite{crolletal11} have presented measurements of the transit-depths in the J- and K$_s$-bands, as well as in a band centered on a methane feature at 1.69~$\mu$m. Their measurements in the J- and K$_s$-band provide evidence for an increase in the planetary radius at 2.15~$\mu$m compared to that at 1.15~$\mu$m. This increase is consistent with the presence of a thick, hydrogen-rich atmosphere. The additional measurement of the transit-depth at 1.69~$\mu$m, which probes a methane feature in the H-band, is lower than that observed in the J-band. This point however has a large uncertainty, and is therefore still consistent with an atmosphere with a low mean-molecular weight, but only if the methane abundance is low. Comparison of these data with our models shows a significant increase in the $\chi^2$ between the sub-solar metallicity model and the water model of $\Delta\chi^2$=20.1, as advocated by~\cite{crolletal11}. The large depth in the K$_s$-band for the sub-solar metallicity model is due to H$_2$-H$_2$ CIA and not due to methane.

In contrast, using the IRAC camera on the {\it Spitzer} Space Telescope,~\cite{desertetal11} find radius ratios in the IRAC bands at 3.6 and 4.5~$\mu$m that are consistent with the observations of~\cite{beanetal10}, showing  a flat transmission spectrum extending into the mid-infrared. The observations by \cite{desertetal11} therefore reject a cloud-free methane rich atmosphere because it would require a significantly larger radius at 3.6$\mu$m than is observed. Within the 1$\sigma$ errorbars, the observations of~\cite{desertetal11} are consistent with both the sub-solar metallicity model with a large scale-height and with the model for a water-dominated atmosphere.

In addition to our own g-band measurement, observations by~\cite{carteretal11} in the r- and z-band show evidence for an increase in the radius-ratio of the planet when going from the z-band to the r-band at the 1$\sigma$ level. However, they also caution that stellar spots are important when comparing observations at different dates and wavelengths. The measured increase of the planet-to-star radius ratio is 0.00184$\pm$0.00166, which is consistent with Rayleigh scattering, where the expected difference is $\sim$0.0015.  It is, however, difficult to directly compare these measurements with our measurements, since~\cite{carteretal11}  used a different impact parameter and semi-major axis.

In the top panel of Figs.~\ref{fig:tspec_all}~and~\ref{fig:tspec_all_opt} we show all the currently available measurements of the transmission spectrum of GJ1214b that were obtained using the same impact parameter and scaled semi-major axis. We caution, however, that when  making a direct comparison between literature values and our observations one has to be careful, since these were taken at different epochs, and therefore at different levels of stellar variability.

Ignoring these subtle effects, the available data favour the model for a hydrogen-rich, low metallicity atmosphere with a cloud layer at 0.5 bar ($\chi^2$=14.3) over a water dominated atmosphere ($\chi^2$=48.7) and a flat line ($\chi^2$=46.8). This is mainly due to our g-band measurements and the measurements in the K$_s$-band presented in this paper and in~\cite{crolletal11}. The {\it Spitzer} measurements from~\cite{desertetal11} are consistent with both models.

\subsection{The impact of unocculted starspots}\label{sec:spotbg}
As discussed in Section~\ref{sec:starvar}, unocculted spots can also produce a wavelength dependent signal in the transmission spectrum, by altering the effective radius of the star. Since only the variation in the starspots is traced by the variability measurements, and not the total level of spots, we have no constraints on the base level of spots present on the stellar surface. There is a priori no reason to assume that the only spots present on the stellar surface are those responsible for the variability. This would result in flat regions in the light curve when no spots are visible on the stellar hemisphere pointed toward Earth, which is not seen in the light curve by~\cite{bertaetal11}.

From this variability curve we know that GJ1214 varies in brightness at the $\sim$1\% level in the MEarth bandpass. This corresponds to a variation in the spot-fraction of $\sim$2.5\%, assuming a spot temperature of 2800K. We might expect that this variability is the result of a distribution of many small spots on the stellar surface. This is supported by the fact that for 2 out of the 16 transits observed by~\cite{carteretal11} the planet crosses over a small star spot. Since these spots are possibly distributed in a random pattern in longitude on the stellar surface, we argue that the fraction of the star that is covered in spots is significantly higher than the 2.5\% necessary to explain the variability. 

 To investigate at what level the observed transmission spectrum can be influenced by a base level of unocculted spots, we calculated the wavelength dependent bias in the radius ratio for several assumed base levels, which is overplotted on the observed transmission spectrum in the middle panels of Figs.~\ref{fig:tspec_all}~and~\ref{fig:tspec_all_opt}. This bias was calculated in the same way as the corrections for stellar variability in Sect.~\ref{sec:starvarcor}. However, instead of the observed stellar variability, we calculated the bias for different amounts of stellar dimming in the GROND i-band. As can be seen it is very unlikely that star spots are responsible for the higher g-band and K$_s$-band values. To explain the g-band measurement, a large part of the stellar surface ($>$20\%) has to be covered in starspots, which would lead to a much lower radius ratio in the near and mid-infrared than observed.

However, this argument can also be turned around. Although the observed transmission spectrum cannot be explained by a baseline of unocculted starspots, if there is a significant baseline present, the shape of the planet's transmission spectrum can also be altered in such a way that the observed transmission spectrum appears relatively flat. We therefore also investigate how the starspots can change the shape of the transmission spectrum. To demonstrate this impact, we applied a correction for the spot base level, assuming that the star is dimmed by 5\% in the i-band, which corresponds to a spot covering fraction of 10\% of the stellar surface. In the bottom panels of Figs.~\ref{fig:tspec_all}~and~\ref{fig:tspec_all_opt}  we show this corrected transmission spectrum, where, to make comparison with the earlier results easier, we have shifted the values such that the i-band radius ratio is unaltered. The effect of this correction is to push the near and mid-infrared points to larger radius ratios with respect to the optical measurements. The increased radius ratio in the infrared is then more consistent with a geometrically thick atmosphere, although we require a strongly reduced methane content in order to explain both our K$_c$-band and IRAC points. At the red part of the optical spectrum the correction introduces a tilt towards smaller radii for shorter wavelenghts, which is opposite to the tilt expected from Rayleigh scattering in the atmosphere of the planet, which causes an increase in the radius ratio. 

 From this example it is clear that not only stellar variability needs to be taken into account when comparing the transmission spectrum at different wavelenghts and times, but that it is also extremely important to investigate the impact of the base level of star spots on the observed transmission spectrum.

\section{Conclusions}\label{sec:concl}
We have presented observations of 8 transits of GJ1214b, in the g-, r-, i-, I- and z-bands in the optical, and K$_s$- and K$_c$-bands in the near-infrared, allowing us to determine the planet-to-star radius ratio of this system at these wavelengths. All measurements show within errors a similar size ratio, except for the g-band, for which this ratio is larger at the $\sim$2$\sigma$ level. If real, this could be attributed to Rayleigh scattering within the planet's atmosphere, and subsequently points to a large scale height and therefore a hydrogen/helium dominated atmosphere. However, this requires more observations to confirm.

When combining our observations with all the currently available literature values, the data again favour a model with an extended atmosphere, a low methane content and a cloud-layer at 0.5 bar, above that for a water dominated atmosphere.

In addition, it is very important to take into account the base-level of stellar spots, which can significantly alter the shape of the transmission spectrum, by increasing the radius ratios measured at optical wavelengths compared to those at near- and mid-infrared wavelengths. When this effect is removed, the relative increases in the planet-to-star size ratios in the near- and mid-infrared make those measurements more consistent with an atmosphere with a large scale-height compared to an atmosphere which is dominated by water vapour.

\subsection*{Acknowledgements}
We are grateful to the staff of the NOT telescope for their assistance with these observations. We like to thank Marco Nardini for assistence with the GROND observations.
Based on observations made with the Nordic Optical Telescope, operated on the island of La Palma jointly by Denmark, Finland, Iceland, Norway, and Sweden, in the Spanish Observatorio del Roque de los Muchachos of the Instituto de Astrofisica de Canarias.
The Isaac Newton Telescope is operated on the island of La Palma by the Isaac Newton Group in the Spanish Observatorio del Roque de los Muchachos of the Instituto de Astrof\'isica de Canarias.
Part of the funding for GROND (both hardware as well as personnel) was generously granted from the Leibniz-Prize to Prof.~G.~Hasinger (DFG grant HA 1850/28-1).

\bibliographystyle{aa} 
\bibliography{demooij_gj1214.bib} 

\begin{thebibliography}{38}
\expandafter\ifx\csname natexlab\endcsname\relax\def\natexlab#1{#1}\fi

\bibitem[{{Acosta-Pulido} {et~al.}(2002){Acosta-Pulido}, {Ballesteros},
  {Barreto}, {Correa}, {Delgado}, {Dominguez-Tagle}, {Hernandez}, {Lopez},
  {Manchado}, {Manescau}, {Moreno}, {Prada}, {Redondo}, {Sanchez}, \&
  {Tenegi}}]{LIRIS02}
{Acosta-Pulido}, J., {Ballesteros}, E., {Barreto}, M., {et~al.} 2002, The
  Newsletter of the Isaac Newton Group of Telescopes, 6, 22

\bibitem[{{Bailey} \& {Kedziora-Chudczer}(2011)}]{baileyandkedziora11}
{Bailey}, J. \& {Kedziora-Chudczer}, L. 2011, ArXiv:1109.3748 [astro-ph.EP]

\bibitem[{{Bean} {et~al.}(2010){Bean}, {Kempton}, \& {Homeier}}]{beanetal10}
{Bean}, J.~L., {Kempton}, E., \& {Homeier}, D. 2010, \nat, 468, 669

\bibitem[{{Berta} {et~al.}(2011){Berta}, {Charbonneau}, {Bean}, {Irwin},
  {Burke}, {D{\'e}sert}, {Nutzman}, \& {Falco}}]{bertaetal11}
{Berta}, Z.~K., {Charbonneau}, D., {Bean}, J., {et~al.} 2011, \apj, 736, 12

\bibitem[{{Borysow}(2002)}]{borysow02}
{Borysow}, A. 2002, \aap, 390, 779

\bibitem[{{Carter} {et~al.}(2011){Carter}, {Winn}, {Holman}, {Fabrycky},
  {Berta}, {Burke}, \& {Nutzman}}]{carteretal11}
{Carter}, J.~A., {Winn}, J.~N., {Holman}, M.~J., {et~al.} 2011, \apj, 730, 82

\bibitem[{{Charbonneau} {et~al.}(2009){Charbonneau}, {Berta}, {Irwin}, {Burke},
  {Nutzman}, {Buchhave}, {Lovis}, {Bonfils}, {Latham}, {Udry}, {Murray-Clay},
  {Holman}, {Falco}, {Winn}, {Queloz}, {Pepe}, {Mayor}, {Delfosse}, \&
  {Forveille}}]{charbonneauetal09}
{Charbonneau}, D., {Berta}, Z.~K., {Irwin}, J., {et~al.} 2009, \nat, 462, 891

\bibitem[{{Charbonneau} {et~al.}(2002){Charbonneau}, {Brown}, {Noyes}, \&
  {Gilliland}}]{charbonneauetal02}
{Charbonneau}, D., {Brown}, T.~M., {Noyes}, R.~W., \& {Gilliland}, R.~L. 2002,
  \apj, 568, 377

\bibitem[{{Claret}(2000)}]{claret00}
{Claret}, A. 2000, \aap, 363, 1081

\bibitem[{{Claret}(2004)}]{claret04}
{Claret}, A. 2004, \aap, 428, 1001

\bibitem[{{Colon} {et~al.}(2010){Colon}, {Ford}, {Redfield}, {Fortney},
  {Shabram}, {Deeg}, \& {Mahadevan}}]{colonetal11}
{Colon}, K.~D., {Ford}, E.~B., {Redfield}, S., {et~al.} 2010, ArXiv:1008.4800
  [astro-ph.EP]

\bibitem[{{Croll} {et~al.}(2011){Croll}, {Albert}, {Jayawardhana},
  {Miller-Ricci Kempton}, {Fortney}, {Murray}, \& {Neilson}}]{crolletal11}
{Croll}, B., {Albert}, L., {Jayawardhana}, R., {et~al.} 2011, \apj, 736, 78

\bibitem[{{de Mooij} {et~al.}(2011){de Mooij}, {de Kok}, {Nefs}, \&
  {Snellen}}]{demooijetal11}
{de Mooij}, E.~J.~W., {de Kok}, R.~J., {Nefs}, S.~V., \& {Snellen}, I.~A.~G.
  2011, \aap, 528, A49+

\bibitem[{{de Mooij} \& {Snellen}(2009)}]{demooijandsnellen09}
{de Mooij}, E.~J.~W. \& {Snellen}, I.~A.~G. 2009, \aap, 493, L35

\bibitem[{{D{\'e}sert} {et~al.}(2011{\natexlab{a}}){D{\'e}sert}, {Bean},
  {Miller-Ricci Kempton}, {Berta}, {Charbonneau}, {Irwin}, {Fortney}, {Burke},
  \& {Nutzman}}]{desertetal11}
{D{\'e}sert}, J.-M., {Bean}, J., {Miller-Ricci Kempton}, E., {et~al.}
  2011{\natexlab{a}}, \apjl, 731, L40+

\bibitem[{{D{\'e}sert} {et~al.}(2009){D{\'e}sert}, {Lecavelier des Etangs},
  {H{\'e}brard}, {Sing}, {Ehrenreich}, {Ferlet}, \&
  {Vidal-Madjar}}]{desertetal09}
{D{\'e}sert}, J.-M., {Lecavelier des Etangs}, A., {H{\'e}brard}, G., {et~al.}
  2009, \apj, 699, 478

\bibitem[{{D{\'e}sert} {et~al.}(2011{\natexlab{b}}){D{\'e}sert}, {Sing},
  {Vidal-Madjar}, {H{\'e}brard}, {Ehrenreich}, {Lecavelier Des Etangs},
  {Parmentier}, {Ferlet}, \& {Henry}}]{desertetal2011}
{D{\'e}sert}, J.-M., {Sing}, D., {Vidal-Madjar}, A., {et~al.}
  2011{\natexlab{b}}, \aap, 526, A12+

\bibitem[{{Gelman} \& {Rubin}(1992)}]{gelmanandrubin92}
{Gelman}, A. \& {Rubin}, D.~B. 1992, Statistical Science, 7, 457–472

\bibitem[{{Gibson} {et~al.}(2011){Gibson}, {Pont}, \& {Aigrain}}]{gibsonetal11}
{Gibson}, N.~P., {Pont}, F., \& {Aigrain}, S. 2011, \mnras, 411, 2199

\bibitem[{{Greiner} {et~al.}(2008){Greiner}, {Bornemann}, {Clemens}, {Deuter},
  {Hasinger}, {Honsberg}, {Huber}, {Huber}, {Krauss}, {Kr{\"u}hler},
  {K{\"u}pc{\"u} Yolda{\c s}}, {Mayer-Hasselwander}, {Mican}, {Primak},
  {Schrey}, {Steiner}, {Szokoly}, {Th{\"o}ne}, {Yolda{\c s}}, {Klose}, {Laux},
  \& {Winkler}}]{greineretal08}
{Greiner}, J., {Bornemann}, W., {Clemens}, C., {et~al.} 2008, \pasp, 120, 405

\bibitem[{{Hauschildt} {et~al.}(1999){Hauschildt}, {Allard}, {Ferguson},
  {Baron}, \& {Alexander}}]{hauschildtetal99}
{Hauschildt}, P.~H., {Allard}, F., {Ferguson}, J., {Baron}, E., \& {Alexander},
  D.~R. 1999, \apj, 525, 871

\bibitem[{{Mandel} \& {Agol}(2002)}]{mandelandagol02}
{Mandel}, K. \& {Agol}, E. 2002, \apjl, 580, L171

\bibitem[{{Markwardt}(2009)}]{markwardt09}
{Markwardt}, C.~B. 2009, in Astronomical Society of the Pacific Conference
  Series, Vol. 411, Astronomical Data Analysis Software and Systems XVIII, ed.
  {D.~A.~Bohlender, D.~Durand, \& P.~Dowler}, 251

\bibitem[{{Miller-Ricci} \& {Fortney}(2010)}]{millerricciandfortney10}
{Miller-Ricci}, E. \& {Fortney}, J.~J. 2010, \apjl, 716, L74

\bibitem[{{Pont} {et~al.}(2008){Pont}, {Knutson}, {Gilliland}, {Moutou}, \&
  {Charbonneau}}]{pontetal08}
{Pont}, F., {Knutson}, H., {Gilliland}, R.~L., {Moutou}, C., \& {Charbonneau},
  D. 2008, \mnras, 385, 109

\bibitem[{{Redfield} {et~al.}(2008){Redfield}, {Endl}, {Cochran}, \&
  {Koesterke}}]{redfieldetal08}
{Redfield}, S., {Endl}, M., {Cochran}, W.~D., \& {Koesterke}, L. 2008, \apjl,
  673, L87

\bibitem[{{Rogers} \& {Seager}(2010)}]{rogersetal10}
{Rogers}, L.~A. \& {Seager}, S. 2010, \apj, 716, 1208

\bibitem[{{Rothman} {et~al.}(2009){Rothman}, {Gordon}, {Barbe}, {Benner},
  {Bernath}, {Birk}, {Boudon}, {Brown}, {Campargue}, {Champion}, {Chance},
  {Coudert}, {Dana}, {Devi}, {Fally}, {Flaud}, {Gamache}, {Goldman},
  {Jacquemart}, {Kleiner}, {Lacome}, {Lafferty}, {Mandin}, {Massie},
  {Mikhailenko}, {Miller}, {Moazzen-Ahmadi}, {Naumenko}, {Nikitin}, {Orphal},
  {Perevalov}, {Perrin}, {Predoi-Cross}, {Rinsland}, {Rotger}, {{\v S}ime{\v
  c}kov{\'a}}, {Smith}, {Sung}, {Tashkun}, {Tennyson}, {Toth}, {Vandaele}, \&
  {Vander Auwera}}]{rothmanetal09}
{Rothman}, L.~S., {Gordon}, I.~E., {Barbe}, A., {et~al.} 2009, \jqsrt, 110, 533

\bibitem[{{Rothman} {et~al.}(2010){Rothman}, {Gordon}, {Barber}, {Dothe},
  {Gamache}, {Goldman}, {Perevalov}, {Tashkun}, \& {Tennyson}}]{rothmanetal10}
{Rothman}, L.~S., {Gordon}, I.~E., {Barber}, R.~J., {et~al.} 2010, \jqsrt, 111,
  2139

\bibitem[{{Sing} {et~al.}(2011{\natexlab{a}}){Sing}, {D{\'e}sert}, {Fortney},
  {Lecavelier Des Etangs}, {Ballester}, {Cepa}, {Ehrenreich},
  {L{\'o}pez-Morales}, {Pont}, {Shabram}, \& {Vidal-Madjar}}]{singetal11a}
{Sing}, D.~K., {D{\'e}sert}, J., {Fortney}, J.~J., {et~al.} 2011{\natexlab{a}},
  \aap, 527, A73+

\bibitem[{{Sing} {et~al.}(2009){Sing}, {D{\'e}sert}, {Lecavelier Des Etangs},
  {Ballester}, {Vidal-Madjar}, {Parmentier}, {Hebrard}, \&
  {Henry}}]{singetal09}
{Sing}, D.~K., {D{\'e}sert}, J.-M., {Lecavelier Des Etangs}, A., {et~al.} 2009,
  \aap, 505, 891

\bibitem[{{Sing} {et~al.}(2011{\natexlab{b}}){Sing}, {Pont}, {Aigrain},
  {Charbonneau}, {D{\'e}sert}, {Gibson}, {Gilliland}, {Hayek}, {Henry},
  {Knutson}, {Lecavelier Des Etangs}, {Mazeh}, \& {Shporer}}]{singetal11b}
{Sing}, D.~K., {Pont}, F., {Aigrain}, S., {et~al.} 2011{\natexlab{b}}, \mnras,
  416, 1443

\bibitem[{{Snellen} {et~al.}(2008){Snellen}, {Albrecht}, {de Mooij}, \& {Le
  Poole}}]{snellenetal08}
{Snellen}, I.~A.~G., {Albrecht}, S., {de Mooij}, E.~J.~W., \& {Le Poole}, R.~S.
  2008, \aap, 487, 357

\bibitem[{{Snellen} {et~al.}(2010){Snellen}, {de Kok}, {de Mooij}, \&
  {Albrecht}}]{snellenetal10}
{Snellen}, I.~A.~G., {de Kok}, R.~J., {de Mooij}, E.~J.~W., \& {Albrecht}, S.
  2010, \nat, 465, 1049

\bibitem[{{Swain} {et~al.}(2008){Swain}, {Vasisht}, \& {Tinetti}}]{swainetal08}
{Swain}, M.~R., {Vasisht}, G., \& {Tinetti}, G. 2008, \nat, 452, 329

\bibitem[{{Tinetti} {et~al.}(2007){Tinetti}, {Vidal-Madjar}, {Liang},
  {Beaulieu}, {Yung}, {Carey}, {Barber}, {Tennyson}, {Ribas}, {Allard},
  {Ballester}, {Sing}, \& {Selsis}}]{tinettietal07}
{Tinetti}, G., {Vidal-Madjar}, A., {Liang}, M., {et~al.} 2007, \nat, 448, 169

\bibitem[{{Vidal-Madjar} {et~al.}(2004){Vidal-Madjar}, {D{\'e}sert},
  {Lecavelier des Etangs}, {H{\'e}brard}, {Ballester}, {Ehrenreich}, {Ferlet},
  {McConnell}, {Mayor}, \& {Parkinson}}]{vidalmadjaretal04}
{Vidal-Madjar}, A., {D{\'e}sert}, J., {Lecavelier des Etangs}, A., {et~al.}
  2004, \apjl, 604, L69

\bibitem[{{Vidal-Madjar} {et~al.}(2003){Vidal-Madjar}, {Lecavelier des Etangs},
  {D{\'e}sert}, {Ballester}, {Ferlet}, {H{\'e}brard}, \&
  {Mayor}}]{vidalmadjaretal03}
{Vidal-Madjar}, A., {Lecavelier des Etangs}, A., {D{\'e}sert}, J., {et~al.}
  2003, \nat, 422, 143

\end{thebibliography}

\end{document}